\newif\ifanonymousversion
\anonymousversionfalse

\documentclass[conference]{IEEEtran}
\IEEEoverridecommandlockouts
\usepackage{cite}
\usepackage{amsmath,amssymb,amsfonts}
\usepackage{algorithmic}
\usepackage{graphicx}
\usepackage{textcomp}
\usepackage{xcolor}
\usepackage{footmisc}
\usepackage{flushend}
\usepackage{soul}
\usepackage{threeparttable}
\usepackage[hyphens]{url}
\usepackage[hyphenbreaks]{breakurl}
\usepackage[hyperindex,breaklinks]{hyperref}
\usepackage{fancyvrb}
\hypersetup{
   breaklinks=true,   % splits links across lines
   colorlinks=true,   % displays links as colored text instead of blocks
   pdfusetitle=true,  % \title and \author values into pdf metadata
}

\newcommand{\email}[1]{\href{mailto:#1}{#1}}

\begin{document}

%%
%% Title
\title{Analyzing ChatGPT's Aptitude in\\ an Introductory Computer Engineering Course}

%%
%% Authors
\ifanonymousversion

\author{Anonymous Submission}

\else

\author{
\IEEEauthorblockN{Sanjay Deshpande}
\IEEEauthorblockA{
\textit{Electrical Engineering}\\
\textit{Yale University}\\
%New Haven, CT, USA\\
\email{sanjay.deshpande@yale.edu}}
\and
\IEEEauthorblockN{Jakub Szefer}
\IEEEauthorblockA{
\textit{Electrical Engineering}\\
\textit{Yale University}\\
%New Haven, CT, USA\\
\email{jakub.szefer@yale.edu}}
}

\fi

\maketitle

\begin{abstract}
ChatGPT has recently gathered attention from the general public and academia as a tool that is able to generate plausible and human-sounding text answers to various questions. One potential use, or abuse, of ChatGPT is in answering various questions or even generating whole essays and research papers in an academic or classroom setting. While recent works have explored the use of ChatGPT in the context of humanities, business school, or medical school, this work explores how ChatGPT performs in the context of an introductory computer engineering course. This work assesses ChatGPT's aptitude in answering quizzes, homework, exam, and laboratory questions in an introductory-level computer engineering course. This work finds that ChatGPT can do well on questions asking about generic concepts. However, predictably, as a text-only tool, it cannot handle questions with diagrams or figures, nor can it generate diagrams and figures. Further, also clearly, the tool cannot do hands-on lab experiments, breadboard assembly, etc., but can generate plausible answers to some laboratory manual questions. One of the key observations presented in this work is that the ChatGPT tool could not be used to pass all components of the course. Nevertheless, it does well on quizzes and short-answer questions. On the other hand, plausible, human-sounding answers could confuse students when generating incorrect but still plausible answers.
\end{abstract}
%\begin{IEEEkeywords}
%component, formatting, style, styling, insert
%\end{IEEEkeywords}
\vspace*{-0.5cm}
\section{Introduction}

Since the launch of ChatGPT (Chat Generative Pre-trained Transformer) on Nov. 30, 2022, a number of research publications and online news articles outline different studies on various uses of ChatGPT in academic settings~\cite{mannie2023chatgpt}. In most studies, researchers have considered the use of ChatGPT on a single exam or test. Researchers have explored how ChatGPT performs on the final exam of a typical MBA core course, such as an Operations Management course. They found it would have received a B to B- grade on the exam~\cite{needleman2023would}. Researchers have further used ChatGPT and found that it can possibly pass the U.S. Medical Licensing Exam~\cite{kung2022performance}. Researchers have also explored ChatGPT’s performance on the Bar exam. ChatGPT was found to earn a passing grade in the Evidence and Torts section of the Multistate Bar Exam~\cite{bommarito2022gpt}. Outside of academia, in a different exam-like setting, ChatGPT was shown to pass Google coding interview~\cite{dreibelbis2023chatgpt} as well.

Existing studies have thoroughly analyzed ChatGPT in settings of different exams or tests, as discussed above. They have, however, not considered the various types of questions and assignments that make up a whole academic course, which entails different types of questions and problems than just answering written exam questions. Further, in computer engineering courses, different questions may require not just writing text or code answers, which ChatGPT can do well, but also making block diagrams, drawing truth tables, etc., which ChatGPT cannot handle well as a text-only tool. In this context, the present work explores how well ChatGPT can do in a computer engineering course, what are the limitations, how it can affect student's learning (or possible cheating), and what improvements could make use (and abuse) of ChatGPT more effective in such a course.

As part of the evaluation, this work introduces a new metric of \textit{accuracy-per-minute}. The metric tries to capture both how good the solutions are and how long it took to generate the solutions. As can be observed from the evaluation presented later in this work, the GPT-3 generated solutions do much better compared to student non-GPT-3 solutions when considering this metric. The key takeaway is that while GPT-3 does worst, or much worst than students in generating correct answers, but it is orders of magnitude faster in generating these~solutions.

\vspace*{-0.1cm}
\subsection{Software Code, Question Files, and Answer Files}

GPLv3 Python code developed and used in this work is available online%
\footnote{
\ifanonymousversion
{\tt URL has been removed for anonymous submission.}
\else
\url{https://github.com/caslab-code/openai-eeng201-solution-generator}
\fi
}. Quiz, homework, exam, laboratory project questions, and respective sample answers generated by the OpenAI API are all available as JSON files in the above repository. Student generated ChatGPT answers to quizzes, homework, exam, and laboratory project solutions are not provided to ensure students' privacy.

\subsection{Disclaimer and ChatGPT Versions used }

ChatGPT was not used in preparation for any part of the text of this manuscript. Further, this work was prepared in Spring 2023 using ChatGPT (which is fine-tuned from a model in GPT-3 series). Current and newer versions of ChatGPT may be more capable. Exploration of the most recent ChatGPT versions is left as future work. 

\section{ChatGPT in an Introductory Computer Engineering Course}

This work explores the use of ChatGPT in EENG 201, ``Introduction to Computer Engineering'' course offered at Yale University every spring semester.  The course has no official prerequisites. It is most often taken by freshmen or sophomores interested in majoring in Electrical Engineering. This may be the first or second course students have taken in an engineering discipline. Many students have some programming background, but this is not required or expected. Further, about $30\%$ of the students in the class typically are more senior Computer Science majors wanting to learn a bit about computer hardware and Computer Engineering. A very small number of students each year also come from Mechanical Engineering or non-engineering fields such as psychology or theater. All students, regardless of their background, generally want to learn about digital design and computer engineering, even if this is not their major.

\subsection{Quizzes, Homeworks, Exams, and Laboratories}

The EENG 201 course uses a number of different student evaluation components:

\begin{itemize} \small
    \item Quizzes -- There are weekly quizzes, where students have $5$ minutes to give written answers to questions presented at the start of class. Questions usually pertain to the prior week's topics discussed in class. Students submit text-only answers online, and the answers contain short written text only.
    \item Homeworks -- There are bi-weekly homework assignments where students have $1$ week to answer questions. Questions are a mixture of textbook questions or questions developed by the instructor or the teaching assistants. Students submit PDF files with answers online; the answers may contain short answers, boolean equations, proofs and simplification of boolean equations, truth tables, K-maps, block diagrams, state machine diagrams, Verilog code, etc.
    \item Exams -- The course contains the usual mid-term and final exams. Exam questions have a similar format to homework questions. Students submit their answers at the end of the exam. In Spring 2023, the exams are open-notes and open-book. Students may use electronic devices to look up course materials or possibly online resources (which they are required to cite if they look up information in non-course related sources).
    \item Laboratory Projects -- The course includes a number of laboratory projects, ranging from implementing a circuit on a breadboard to writing Verilog code and prototyping it on an FPGA board. Each laboratory manual typically has a number of questions that students must answer as they work their way through the manual. At the end of each laboratory, students also need to demonstrate their working project.
\end{itemize}

\subsection{Students Prior Experiences with ChatGPT}

An informal vote at the beginning of the course in January 2023 indicated that most students know about ChatGPT. Roughly 75\% have indicated they are familiar with ChatGPT and have used it. Informally, some students discussed using ChatGPT to help with essays or written work in other courses. Thus students are well aware of ChatGPT and generally have experimented with this technology prior to the class.

\subsection{Student ChatGPT-assisted Solutions}

In Spring 2023, students were offered an option to use ChatGPT in EENG 201 following a set of rules. An abbreviated list of rules is summarized below:

\begin{itemize} \small
    \item Students must submit main, graded written work (quiz, assignment, exam, etc.) without the use of ChatGPT.
    \item Students can submit a second version of their work where answers were generated by ChatGPT.
    \item The ChatGPT-assisted work will be graded to assess how well ChatGPT does, but this will not affect students'~grades.
    \item Students who submit the second, ChatGPT assisted solutions can recover up to 2\% of the grade on the corresponding written work (quiz, assignment, exam, etc.).
\end{itemize}

\noindent The last item listed above was incorporated to give students an incentive not to try to cheat by using ChatGPT on the main work while still giving them the motivation to take part in the experiment of using ChatGPT in the class.

\subsection{OpenAI API Generated Solutions}

In parallel to students using ChatGPT to answer questions, a set of scripts was developed by us to submit class-related questions to the OpenAI API automatically and to generate answers to the questions automatically. Questions were transcripted into JSON files for all quizzes, assignments, exams, etc. The scripts then used the JSON files to query the OpenAI API and generate solutions. Students did not have access to this code.

As a result, we can compare student generated ChatGPT answers with automatically generated answers using OpenAI API. In both cases, the GPT-3 model was used, but as demonstrated below, the results and scores are somewhat different based on whether ChatGPT was used (through the web interface) or if OpenAI API was used through the scripts.

\section{Setup for Evaluation of ChatGPT in the Course}

The ChatGPT was evaluated on almost all aspects of the course: quizzes, homework, exams, and laboratory experiments. In particular, the work evaluated includes:

\begin{itemize} \small
    \item $6$ in-class quizzes,
    \item $3$ bi-weekly homework assignments,
    \item $1$ mid-term exam,
    \item $4$ laboratory projects, and
    \item $1$ laboratory practical exam.
\end{itemize}

The current version of the manuscript excludes the mid-term exam and laboratory practical, as these were either not conducted or not graded as of the writing of this article but will be included in the updated version of the article.

\subsection{Questions used in Quizzes, Homeworks, Exams, and Laboratory Projects}

To attempt to limit bias when developing questions, i.e., try not to intentionally make questions too difficult or too easy for ChatGPT, the number of questions were re-used from the prior year's quizzes and assignments when possible. This way, the old questions were guaranteed to have been developed without knowledge that ChatGPT would be used. Further, questions from the textbook were clearly developed before ChatGPT, as the course textbook%\footnote{\url{https://dl.acm.org/doi/10.5555/2381028}} 
was published in 2012. Most of the work, i.e., quizzes, homework, and exams, included fully new questions for Spring 2023 as well. There may be unconscious bias in these questions, but as much as possible, care was taken to develop ``typical'' questions regardless of the knowledge that ChatGPT will be used to try to answer them.

As one point of note, by the nature of the quizzes (text-only short answers or coding questions), they are easier to answer by only writing text, compared to homework or exam questions that require diagrams, for example. Homeworks and exams are harder to answer by writing text, as some questions reference a diagram or a figure or ask students to make a diagram or a figure. Table~\ref{type_of_questions} shows the percentage of questions that are text-based versus figure-based in each assigned set of questions. Laboratory project questions typically ask for simplification of some boolean equations or the development of Verilog code. Still, hands-on demonstrations or assembly of breadboards, for example, is clearly out of reach of text-based ChatGPT.

\begin{table}[t]
\caption{\scriptsize {Percentage of Text-based figure-based questions in the assigned work.$^*$}}
\begin{center}
\scriptsize
\begin{tabular}{|l|c|c|}
\hline
\textbf{Assigned Work} & \multicolumn{1}{l|}{\textbf{\% Text-based}} & \multicolumn{1}{l|}{\textbf{\% Fig-based }} \\ \hline
Quiz, Week 2           & 100.00\%                                              & 0.00\%                                               \\ \hline
Quiz, Week 3           & 100.00\%                                              & 0.00\%                                               \\ \hline
Quiz, Week 4           & 100.00\%                                              & 0.00\%                                               \\ \hline
Quiz, Week 5           & 100.00\%                                              & 0.00\%                                               \\ \hline
Quiz, Week 6           & 100.00\%                                              & 0.00\%                                               \\ \hline
Quiz, Week 7           & 100.00\%                                              & 0.00\%                                               \\ \hline
Assignment 1           & 58.33\%                                               & 41.67\%                                              \\ \hline
Assignment 2           & 32.08\%                                               & 67.92\%                                              \\ \hline
Assignment 3           & 27.36\%                                               & 72.64\%                                              \\ \hline
Mid-term Exam          & ---                                                    & ---                                \\ \hline
Lab 1 Questions        & 66.67\%                                               & 33.33\%                                              \\ \hline
Lab 2 Questions        & 75.00\%                                               & 25.00\%                                              \\ \hline
Lab 3 Questions        & 100.00\%                                              & 0.00\%                                               \\ \hline
Lab 4 Questions        & 100.00\%                                              & 0.00\%                                               \\ \hline
Lab Practical exam     & ---                                                    & ---                               \\ \hline
\end{tabular}
\label{type_of_questions}
\end{center}
% \begin{minipage}{\linewidth} 
\tiny{$^*$Here \textit{Text-based} questions represent textual question input and textual expected output. In contrast, \textit{Figure-based} questions represent either the question containing a figure or the expected answer containing a figure or both.}
% \end{minipage}
\vspace{-7mm}
\end{table}

\subsection{Grading Approach}

All the work in the course was graded by the instructor or the teaching assistants. Most works was graded using an online system where text files or PDF files of the answers can be viewed, marked with notes, and graded. Student solutions generated with ChatGPT were graded this way as well through the same online system. OpenAI API generated solutions were automatically converted to human-friendly PDF files for grading. They were graded offline as there was no way to upload these PDFs to the class grading system.

\subsection{Student Submitted Answers using ChatGPT}

About $28\%$ of students chose to submit ChatGPT assisted work in addition to their main work. Not all students submitted answers for all types of work. A small number of students chose to only take part in the ChatGPT experiment for quizzes, for example. The reported scores are averages of the work submitted by the students.

The work was graded by the instructor and teaching assistant. The teaching assistant was instructed, to the best of their abilities, to grade the ChatGPT-assisted work as if it were student work. The teaching assistant was informed there is no preset conclusion that we want to arrive at, and thus to try to grade as objectively as possible. Due to this setup of the grading system, the teaching assistant did know that this was ChatGPT assisted work. There could be unconscious bias in the grading due to this aspect.

\subsection{Automated Answers Generated using OpenAI API}

In addition to the student work, OpenAI API and Python scripts were used to generate answers to questions. Three sets of answers were generated for each type of work, so the reported averages for solutions generated using OpenAI API are averages of three solutions.

For each quiz, homework, exam, and laboratory project, a JSON data file containing text and figures for any of the questions was prepared. The questions were the same ones presented to the students. Textbook questions were transcribed into text format. Very minor modifications were performed on the questions when saving them as JSON files. Typical modifications included separating questions with sub-questions into separate questions, e.g., questions with $3$ sub-parts would be changed into $3$ separate questions in the JSON file, each submitted separately to the API. Some text would have to be repeated when sub-parts were changed into separate questions. The JSON files, and the Python scripts discussed later, are made available online by the authors.%
\footnote{
\label{footnote1}JSON files and GPLv3 code developed for use in this work are available online at
\ifanonymousversion
{\tt URL has been removed for anonymous submission}.
\else
\url{https://github.com/caslab-code/openai-eeng201-solution-generator}.
\fi
}

To generate answers to the questions, Python scripts were written to query OpenAI API and to obtain answers for each question. Clearly, OpenAI API does not handle image files, so only text, but not images, was submitted for each question. The answers were saved as JSON files as well. A set of $3$ JSON files with answers for each written work was generated. The generated answer files in the JSON format are also available online.%
\footref{footnote1}
Further, scripts were used to generate more human-friendly PDF files from the JSON files, which can be used for~grading.

These generated answers in the PDF format were likewise graded by the instructor and teaching assistant, just as the student generated work. Since it is clear that ChatGPT generated JSON files are being graded, there may be unconscious bias in the grading. Further, students would typically use ChatGPT chat, which maintains some context, so the answers may be better than the API. To the author's best knowledge, the API calls do not maintain context, so it seems that answers generated using the API may be less accurate given the lack of context about prior questions asked. More formally, the OpenAI API was used in a zero-shot learning approach, while online ChatGPT is able to maintain context between questions being asked.

\subsection{OpenAI Completion API Configuration}

To generate the answers automatically using a script, OpenAI API's Completion feature was used. It creates a completion for the provided prompt and parameters. The Completion configuration used is listed below. The {\tt text} is the answer or solution text, while the {\tt prompt} is the input question:

\begin{Verbatim}[fontsize=\small]
     text = openai.Completion.create(
       model="text-davinci-003",
       prompt=prompt,
       max_tokens=256,
       temperature=1,
       n=1)    
\end{Verbatim}

\section{Evaluation Results}

Table~\ref{comb_tab} {shows the average scores for the student solutions which were written by students without any help of ChatGPT, and the average scores for the student solutions which were generated with the help of ChatGPT. The table shows that quiz answers generated by ChatGPT can reach $100\%$. Assignment solutions generated using ChatGPT were the least accurate, often below passing grade. Solutions relating to laboratory project questions were in-between.

\begin{table*}[t]
\caption{\scriptsize {Evaluation of solutions written by students without any assistance of ChatGPT, solution generated by students with assistance of ChatGPT, and solutions generated using python scripts and OpenAI API.}}
\scriptsize
\begin{center}
\begin{tabular}{|c|c|c|c|c|c|c|c|c|c|}
\hline
\textbf{Assigned}&\multicolumn{3}{|c|}{\textbf{Student Solutions without ChatGPT}}&\multicolumn{3}{|c|}{\textbf{Student ChatGPT-assisted Solutions}}&\multicolumn{3}{|c|}{\textbf{OpenAI API-generated Solutions}} \\
\cline{2-10} 
\textbf{Work} & \textbf{\textit{Avr. Score}}& \textbf{\textit{Max. Points}}& \textbf{\textit{Percentage}} & \textbf{\textit{Avr. Score}}& \textbf{\textit{Max. Points}}& \textbf{\textit{Percentage}} & \textbf{\textit{Avr. Score}}& \textbf{\textit{Max. Points}}& \textbf{\textit{Percentage}} \\
\hline
\hline
Quiz, Week 2 & $2.7$ & $3$ & $89\%$    & $3.0$ & $3$ & $100\%$      & $1.8$ & $3$ & $60\%$           \\\hline
Quiz, Week 3 & $2.8$ & $3$ & $94\%$    & $2.9$ & $3$ & $97\%$       & $2.5$ & $3$ & $83\%$           \\\hline
Quiz, Week 4 & $2.6$ & $3$ & $86\%$    & $2.6$ & $3$ & $87\%$       & $1.8$ & $3$ & $60\%$           \\\hline
Quiz, Week 5 & $2.5$ & $3$ & $84\%$    & $3.0$ & $3$ & $100\%$      & $1.3$ & $3$ & $43\%$           \\\hline
Quiz, Week 6 & $2.7$ & $3$ & $89\%$    & $2.7$ & $3$ & $90\%$       & $2.5$ & $3$ & $83\%$           \\\hline
Quiz, Week 7 & $2.9$ & $3$ & $95\%$    & $3.0$ & $3$ & $100\%$      & $2.2$ & $3$ & $73\%$           \\\hline \hline
Assignment 1 & $22.2$  & $25$ & $89\%$ & $6.92$  & $25$ & $27\%$    & $9.2$ & $25$ & $37\%$           \\\hline
Assignment 2 & $22.9$  & $25$ & $91\%$ & $1.67$  & $25$ & $8\%$     & $1.9$ & $25$ & $8\%$           \\\hline
Assignment 3 & $21.7$ & $25$ & $87\%$  & $3.00$ & $25$ & $13\%$     & $2.0$ & $25$ & $8\%$           \\\hline\hline
Mid-Term Exam & --- & $50$ & ---       & --- & $50$ & ---           & --- & $50$ & ---              \\\hline\hline
Lab 1 Questions & $5$ & $5$ & $100\%$  & $2.90$ & $5$ & 64\%        & $2.2$ & $5$ & $44\%$        \\\hline
Lab 2 Questions & $5$ & $5$ & $100\%$  & $2.70$ & $5$ & 60\%        & $2.8$ & $5$ & $56\%$        \\\hline
Lab 3 Questions & $5$ & $5$ & $100\%$  & $3.00$ & $5$ & 70\%        & $2.8$ & $5$ & $56\%$        \\\hline
Lab 4 Questions & $5$ & $5$ & $100\%$  & $2.88$ & $5$ & 60\%        & $2.2$ & $5$ & $44\%$        \\\hline\hline
Lab Practical Exam & --- & $30$ & ---  & --- & $30$ & ---           & --- & $30$ & ---         \\\hline
\end{tabular}
\label{comb_tab}
\end{center}
\vspace{-8mm}
\end{table*}

Table~\ref{comb_tab} also shows the average scores of the solutions generated using automated scripts. Here quiz solution accuracy was less, but still above passing grade for most quizzes. Assignment solutions generated with the OpenAI API were somewhat similar to ChatGPT solutions but still below passing grades. Solutions relating to laboratory project questions were in-between.

Table~\ref{tab3} shows the approximate time taken to generate solutions. Students self-reported the time taken to generate the solutions, with the exception of quizzes and mid-term exams where they were fixed to $5$ min. and $75$ min. respectively. As of the writing of this article, the collection of self-reported time to generate solutions for some work is ongoing and corresponding entries are labeled with --- in the table. It can be seen that the automated approach of using OpenAI API to generate the solutions is orders of magnitude faster than students self-generating and writing solutions.

\begin{table}[t]
\caption{\scriptsize Approximate time to complete the work.$^\dagger$}
\begin{center}
\scriptsize
\begin{tabular}{|c|c|c|c|}
\hline
\textbf{}&\multicolumn{3}{|c|}{\textbf{Approx. Time to Generate Solutions}} \\
\cline{2-4} 
\textbf{Assigned} & \textbf{\textit{Student}} & \textbf{\textit{Student}} & \textbf{\textit{Automated}} \\
\textbf{Work} & \textbf{\textit{no ChatGPT}} & \textbf{\textit{with ChatGPT}} & \textbf{\textit{Code}} \\
\textbf{} & \textbf{\textit{(min.)}} & \textbf{\textit{(min.)}} & \textbf{\textit{(min.)}} \\
\hline
\hline
Quiz, Week 2 & $5$ & --- & $0.3$ \\
\hline
Quiz, Week 3 & $5$ & --- & $0.4$ \\
\hline
Quiz, Week 4 & $5$ & --- & $0.2$ \\
\hline
Quiz, Week 5 & $5$ & --- & $0.4$ \\
\hline
Quiz, Week 6 & $5$ & --- & $0.3$ \\
\hline
Quiz, Week 7 & $5$ & --- & $0.3$ \\
\hline
\hline
Assignment 1 & $380$ & --- & $3.0$ \\
\hline
Assignment 2 & $560$ & --- & $2.3$ \\
\hline
Assignment 3 & $800$ & --- & $1.7$ \\
\hline
\hline
Mid-Term Exam & $75$ & --- & $1.0$ \\
\hline
\hline
Lab 1 Questions & $68$ & --- & $0.3$ \\
\hline
Lab 2 Questions & $72$ & --- & $0.5$ \\
\hline
Lab 3 Questions & $79$ & --- & $0.4$ \\
\hline
Lab 4 Questions & $73$ & --- & $0.4$ \\
\hline
\hline
Lab Practical Exam & $86$ & --- & $0.4$ \\
\hline
\end{tabular}
\label{tab3}
\end{center}
% \begin{minipage}{\linewidth} 
\tiny{$^\dagger$Except For quizzes, which have fixed time of 5 min., and mid-term exam, which has fixed time of 75 min., average of the student self-reported time is reported in the table. For all solutions generated using Python scripts, average time reported by the scripts is given; majority of this time is spent querying the OpenAI API. Lower value is better.}
% \end{minipage}
\vspace{-6mm}
\end{table}

Table~\ref{tab4} evaluates the new metric of accuracy-per-minute. The metric tries to capture both how good the solutions are and how long it took to generate the solutions. Examining the table it can be seen that the GPT-3 generated solutions do much better when considering this metric: while GPT-3 does worst, or much worst than students in generating correct answers, it is orders of magnitude faster in generating the solutions. Thus the worst accuracy of the solutions is overpowered by the much faster generation time.

\begin{table}[t]
\caption{\scriptsize Efficiency of generating the solutions, in percentage per minute. Higher value is better.}
\begin{center}
\scriptsize
\begin{tabular}{|c|c|c|c|}
\hline
\textbf{}&\multicolumn{3}{|c|}{\textbf{Solution Generation Efficiency}} \\
\cline{2-4} 
\textbf{Assigned} & \textbf{\textit{Student}} & \textbf{\textit{Student}} & \textbf{\textit{Automated}} \\
\textbf{Work} & \textbf{\textit{no ChatGPT}} & \textbf{\textit{with ChatGPT}} & \textbf{\textit{Code}} \\
\textbf{} & \textbf{\textit{(\% / min.)}} & \textbf{\textit{(\% / min.)}} & \textbf{\textit{(\% / min.)}} \\
\hline
\hline
Quiz, Week 2 & $17.8$ & --- & $200$ \\
\hline
Quiz, Week 3 & $18.8$ & --- & $208$ \\
\hline
Quiz, Week 4 & $17.2$ & --- & $300$ \\
\hline
Quiz, Week 5 & $16.8$ & --- & $108$ \\
\hline
Quiz, Week 6 & $17.8$ & --- & $277$ \\
\hline
Quiz, Week 7 & $19.0$ & --- & $243$ \\
\hline
\hline
Assignment 1 & $0.23$ & --- & $12$ \\
\hline
Assignment 2 & $0.16$ & --- & $3.5$ \\
\hline
Assignment 3 & $0.11$ & --- & $4.7$ \\
\hline
\hline
Mid-Term Exam & --- & --- & --- \\
\hline
\hline
Lab 1 Questions & $1.47$ & --- & $146$ \\
\hline
Lab 2 Questions & $1.39$ & --- & $112$ \\
\hline
Lab 3 Questions & $1.27$ & --- & $140$ \\
\hline
Lab 4 Questions & $1.37$ & --- & $110$ \\
\hline
\hline
Lab Practical Exam & --- & --- & --- \\
\hline
\end{tabular}
\label{tab4}
\end{center}
\vspace{-7mm}
\end{table}

\section{Discussion}

This section briefly discusses observations and experiences of using ChatGPT and OpenAI API in an introductory computer engineering course.

\subsection{ChatGPT vs. OpenAI API}

The main observation gained during this work was that web-based ChatGPT seems to do better than the OpenAI API. The difference is presumably that with OpenAI API, the API currently does not keep a context of the questions. Meanwhile, ChatGPT has the ability to retain some context.

A sample example of this is that when asked about ``bubble pushing'' (a visual method presented in the textbook for performing DeMorgan's law), ChatGPT does better because, from prior questions, it may compute that bubble pushing is being used in the context of digital design. Meanwhile, with an OpenAI API call, the API may generate responses about ``bubble pushing'' being some form of market manipulation or a software engineering method.

\subsection{Observation About Impact of API Parameters}

When {\tt max\_tokens} variable is too small, some answers get cut off mid-sentence. The system does not gracefully generate shorter answers based on a smaller {\tt max\_tokens}, but simply cuts off the generated text when {\tt max\_tokens} is reached. Resubmitting the same question with higher {\tt max\_tokens} values can allow for the generation of a full answer. Examples of questions that get cut off include questions asking for truth tables, where the API generates an ASCII text of a possibly very long truth table.

\subsection{Other Observations}

The API sometimes outputs answers which include what looks like ASCII art depicting transistor, gate, or block diagrams. However, they are mostly non-readable.

The API also sometimes outputs links to images. Some links are actually active (often taking users to an image hosting web page), but many do not include a full URL or may not be opened. ChatGPT's web-based interface seems to perform better in terms of displaying images.

\subsection{Cost of Using ChatGPT and OpenAI API}

ChatGPT is available for free online.  OpenAI API was also available for free, or there is a pay-as-you-go version. The pay-as-you-go version was used for this work as it allows for more reliable access with less downtime.

The total cost spent on OpenAI API calls was about \$$2.14$. The work used the OpenAI API GPT-3 Davinci model, which, as of this article's writing, costs \$$0.1200$ per $1$K tokens. According to OpenAI, $1$K tokens are about $750$ words. Other models can be less expensive but may be less accurate. They were currently not tried.

\subsection{Possible Improvements}

The main takeaway about improvements is that there would be a need for ChatGPT, or a supplementary tool, to handle images. Specifically, there are many AI models for labeling images of animals, detecting traffic signs, etc. However, in the context of a computer engineering course, the need is to directly analyze an image for its contents. This may be closer to optical character recognition (OCR) than to AI image inference. Some of the key tasks that would be needed include:

\begin{itemize} \small
    \item Recognize truth tables and the contents of truth tables to determine what logic function it represents and generate Boolean equations from the truth tables.
    \item Recognize transistor diagrams and interpret the connection of the transistors and, thus, the resulting functionality of the gate.
    \item Recognize logic gate diagrams and what combinatorial logic they represent
    \item Recognize finite state machine drawings and what functionality is implemented by the state machine.
    \item Recognize waveform diagrams and then compute how a piece of logic, e.g., a latch or flip-flop, would respond to the inputs shown in the waveform diagrams.
\end{itemize}

Further, in addition to recognizing diagrams and figures discussed above, a needed improvement would be to generate them. I.e., the ability to generate truth tables, transistor diagrams,~etc.

In addition, the multiple tasks may be combined. That is: First, recognize a diagram or the contents of the diagram. Second, perform some computation or simplification based on the information. Third, generate new or updated figures or diagrams.

\section{Conclusion}

This work explored the use of how effective GPT-3 could be in generating solutions to various types of quizzes, assignments, laboratory project questions, and exams in an introductory computer engineering course. This work found that using both the ChatGPT web interface and OpenAI API, correct solutions to quiz questions can often be generated. On the other hand, solutions to homework questions were much less accurate. Many of the limitations can be attributed to the fact that introductory computer engineering coursework requires not just the generation of text or code but also working with figures, block diagrams, state machine diagrams, etc. In its current form, ChatGPT or OpenAI API could not be used to pass an introductory computer engineering course. It is also not likely to help students (neither to learn the material in the course nor to cheat in the course). However, if a ChatGPT-like solution can be developed that is able to process or generate figures, block diagrams, state machine diagrams, etc., it would have a much better chance of passing an introductory computer engineering course.

\bibliographystyle{plain}
\bibliography{refs}

\begin{thebibliography}{1}

\bibitem{bommarito2022gpt}
Michael Bommarito~II and Daniel~Martin Katz.
\newblock Gpt takes the bar exam.
\newblock {\em arXiv preprint arXiv:2212.14402}, 2022.

\bibitem{dreibelbis2023chatgpt}
Emily Dreibelbis.
\newblock {C}hat{G}{P}{T} {P}asses {G}oogle {C}oding {I}nterview for {L}evel 3
  {E}ngineer {W}ith \$183{K} {S}alary --- pcmag.com.
\newblock
  \url{https://www.pcmag.com/news/chatgpt-passes-google-coding-interview-for-level-3-engineer-with-183k-salary},
  2023.
\newblock [Accessed 12-Feb-2023].

\bibitem{kung2022performance}
Tiffany~H Kung, Morgan Cheatham, Arielle Medinilla, ChatGPT, Czarina Sillos,
  Lorie De~Leon, Camille Elepano, Marie Madriaga, Rimel Aggabao, Giezel
  Diaz-Candido, et~al.
\newblock Performance of chatgpt on usmle: Potential for ai-assisted medical
  education using large language models.
\newblock {\em medRxiv}, pages 2022--12, 2022.

\bibitem{mannie2023chatgpt}
Kathryn Mannie.
\newblock {C}hat{G}{P}{T} passes exams for {M}{B}{A} courses and medical
  licences — and it’s only getting started - {N}ational | {G}lobalnews.ca
  --- globalnews.ca.
\newblock
  \url{https://globalnews.ca/news/9432503/chatgpt-exams-passing-mba-medical-licence-bar/},
  2023.
\newblock [Accessed 12-Feb-2023].

\bibitem{needleman2023would}
Emma Needleman.
\newblock {W}ould {C}hat {G}{P}{T} {G}et a {W}harton {M}{B}{A}? {N}ew {W}hite
  {P}aper {B}y {C}hristian {T}erwiesch --- mackinstitute.wharton.upenn.edu.
\newblock
  \url{https://mackinstitute.wharton.upenn.edu/2023/would-chat-gpt3-get-a-wharton-mba-new-white-paper-by-christian-terwiesch/},
  2023.
\newblock [Accessed 12-Feb-2023].

\end{thebibliography}

\end{document}